\documentclass[12pt]{article}
\usepackage{latexsym,amsmath,amssymb,}
\usepackage{graphicx}

\def\be{\begin{equation}}
\def\ee{\end{equation}}
\def\beq{\begin{eqnarray}}
\def\eeq{\end{eqnarray}}

\def\bes{\begin{eqnarray}}
\def\ees{\end{eqnarray}}

\newlength{\sizeonefig}
\newlength{\sizetwofig}
\begin{document}

\title{Connection between black-hole quasinormal modes and lensing in the strong deflection limit}

\author{Ivan Zh. Stefanov$^{1}$\thanks{E-mail: izhivkov@yahoo.com}\,\,\,, \,\, Stoytcho S. Yazadjiev$^{2}$\thanks{E-mail: yazad@phys.uni-sofia.bg}\,\,\,, \,\,
Galin G. Gyulchev$^{3}$\thanks{E-mail: gyulchev@phys.uni-sofia.bg}\\\\
{\footnotesize{$^{1}$ Department of Applied Physics, Technical University of Sofia}}\\
{\footnotesize 8, St. Kliment Ohridski Blvd., 1000 Sofia, Bulgaria}\\\\[-3.mm]
{\footnotesize $^{2}$ Department of Theoretical Physics, Faculty of Physics, St.Kliment Ohridski University of Sofia}\\
{\footnotesize  5, James Bourchier Blvd., 1164 Sofia, Bulgaria }\\\\[-3.mm]
{\footnotesize $^{3}$ Department of Physics, Biophysics and Roentgenology, }\\
{\footnotesize  Faculty of Medicine, St.Kliment Ohridski University of Sofia}\\
{\footnotesize  1, Kozyak Str., 1407 Sofia, Bulgaria}}

%\date{}

%\maketitle

\maketitle

\begin{abstract}
The purpose of the current Letter is to give some relations between gravitational lensing in the strong-deflection limit and the frequencies of the quasinormal modes of spherically symmetric, asymptotically flat black holes. On the one side, the obtained relations can give à physical interpretation of the strong-deflection limit parameters. On the other side, they also
give an alternative method for the measurement of the frequencies of the quasinormal modes of spherically symmetric, asymptotically flat black
holes. They could be applied to the localization of the sources of gravitational waves and could tell us what frequencies of the gravitational waves we could expect from a black hole acting simultaneously as a gravitational lens and a source of gravitational waves.

%Pacs: 04.70.Bw,04.50.Gh,05.45.-a
\end{abstract}

%\pacs{04.70.Bw,04.50.Gh,05.45.-a}

\maketitle

%%%%%%%%%%%%%%%%%%%%%%%%%%%%%%%%%%%%%%%%%%%%%%%%%%%%%%%%%%%%%%%%%%%
Black holes are  extreme objects that are predicted to occur in strong fields by many gravitational theories including general relativity (GR). The
observation of astrophysical objects with enormous masses gives hope that black holes really exist. But for a better determination of the particular
properties of these objects and for a more definitive answer to the question if they are really black holes a more thorough study is needed.

%On the one side,
Information about the physical properties of different compact objects can be obtained through the emission of gravitational waves and the characteristic frequencies of ringing, namely the frequencies of the quasinormal modes (QNMs) \cite{Kokkotas:1999bd, Nollert:1999ji,cardosotopical}. The QNMs would allow us to distinguish between different compact objects -- black holes, stars, to distinguish between different theories of gravity since they predict different characteristic spectra, to determine the global asymptotic charges like mass, charge and angular momentum of the observed black holes and so on \cite{Dreyer}.

The usual method used to calculate the QNM frequencies is to consider a classical scattering problem in a black-hole spacetime. The QNMs correspond to the resonances of the scattering problem when at spacial infinity there are purely outgoing waves and at the event horizon purely ingoing waves. There are also alternative approaches. In the geometric-optics (eikonal) limit Mashhoon \cite{mashhoon} suggested an analytical method for the calculation of the QNMs.  In that approximation gravitational waves are treated as massless particles propagating along the last null unstable, circular orbit (An alternative approach, based on the complex angular momentum method, for the study of resonant scattering in black hole physics is to  interpret the gravitational radiation emitted from black holes in many processes as surface waves localized close to the  last null unstable, circular orbit \cite{Andersson, Folacci1, Folacci2, Folacci3}.  ) and slowly leaking out to infinity \cite{pressringdown} -- \cite{Berti:2005eb}. The real part of the QNMs can be related to the angular velocity of the last null circular orbit and the imaginary part is related to the Lyapunov exponent that determines the instability time scale of the orbit \cite{Cardoso}.

The null geodesics are related also to the propagation of light rays and in the case when the light rays are restricted to a plane the last null
unstable, circular orbit is simply the intersection of the photon sphere with the plane of propagation. Hence, it is natural to
expect the presence of the connection between the two phenomena -- the gravitational
lensing (For a nice, recent and pedagogical review on black-hole gravitational lensing we refer the reader to \cite{Bozza_review}. There one can also find a comprehensive discussion on the observational perspectives.)
and emission of gravitational waves. A guess for a possible connection between these two phenomena has been previously made by
Decanini and Folacci \cite{Folacci2}, though from a different aspect.
%  in a different context. On the bases of that connection they predict possible gravitational lensing observational consequences coming from dispersion relations for the photons such as time delay and hyperfine structure in the system of images.}
In the current Letter some simple
relations between QNMs and the gravitational lensing are presented and  a method for the measurement of the frequencies of the QNMs of
spherically symmetric black holes through gravitational lensing is proposed.

In \cite{Bozza1} Bozza gives a
detailed derivation of the equation for the deflection angle for the case when the lens is
spherically symmetric. He uses the following ansatz for the line element of a generic spherically symmetric spacetime
\cite{Bozza1}
\begin{equation}
ds^2=A(x)dt^2-B(x)dx^2-C(x)\left( d\theta^2+\sin^2\theta d\varphi^2
\right).\label{metric}
\end{equation}
where the metric functions should have the proper asymptotics
\begin{equation}
\begin{array}{c}
  A(x) \longrightarrow^{\hspace{-.7cm} x\rightarrow \infty} 1 - \frac{2M}{x}
  \\ \\
  B(x) \longrightarrow^{\hspace{-.7cm} x\rightarrow \infty} 1 +
  \frac{2M}{x} \\ \\
  C(x) \longrightarrow^{\hspace{-.7cm} x\rightarrow \infty} x^2,
\end{array}
\end{equation}
in order to correctly match the weak gravitational field far from
the lensing object.

The spherical symmetry allows us to restrict our considerations to propagation of light rays in a plane -- say the equatorial plane. Then on the last null circular orbit the following relation holds
\begin{equation}
\frac{C'(x)}{C(x)}=\frac{ A'(x)}{A(x)} \label{Eqm}.
\end{equation}
In the general case Equation (\ref{Eqm}) admits more than one solution. The largest, positive root of Eq. (\ref{Eqm}) gives the radius of the photon
sphere $x_m$. (Here and below we will use the subscript \emph{m} to denote quantities that are evaluated on the photon sphere. For
a more general definition of the photon sphere, we refer the reader to \cite{CVE}).
$A$, $B$, $C$, $A'$ and $C'$ must be positive for $x>x_m$. When restricted to the equatorial plane the photon sphere gives the last circular, null
orbit, which is unstable.

Bozza obtains the following equation for the deflection angle
\begin{equation}
\alpha(\theta)=-\overline{a} \ln \left( \frac{\theta
D_{OL}}{u_m} -1
\right) +\overline{b}, \label{alpha theta}\\%
\end{equation}
where $\theta$ is the angular position of the light source, $\alpha$ is the deflection angle and $D_{OL}$ is the distance between the observer and the
lens. The coefficients $\overline{a}$,  $\overline{b}$ and the impact parameter $u_m$ can be expressed with the metric functions evaluated on the photon sphere in the
following way

\begin{eqnarray}
 \overline{a}&=& \sqrt{{2A_m B_m\over C''_mA_m-C_m A_m''}}\label{a}\\%
 \overline{b}&=&-\pi+b_R+\nonumber\\&&+\overline{a}\ln\left({\frac{ C_m \left( 1- A_m \right)^2
\left(C''_m A_m-C_m A''_m \right)}{A_m^3 {C'_m}^2}}\right),\label{b}\\
u_m&=&\sqrt{\frac{C_m}{A_m}}. \label{um}
\end{eqnarray}
The parameter $b_R$ is different for different space-times and can be obtained through a straightforward calculation. The expression for $b_R$ can be found in \cite{Bozza1}. Prime $(..)'$ denotes the derivative with respect to the radial coordinate $x$.
The strong field lens parameters $\overline{a}$ and $\overline{b}$ carry information about the nature of the lens. Observational data for them could
allow us to distinguish between different black holes, for example a Schwarzschild and a Reissner-Nordstr\"{o}m black hole. (For more information we
refer the reader to \cite{Bozza1}, for example.) The physical meaning of $\overline{a}$ and $\overline{b}$, however, remains somehow obscure.

Let us now consider the method for the calculation of the frequencies of the QNMs through the study of null geodesics. In the eikonal (subdominant corrections to the eikonal approximation can be found in \cite{Dolan}) limit they can be
related to the parameters of the last circular, null geodesic in the following way:
\be\label{QNMLyapunov}
\omega_{\rm QNM}=\Omega_m\,l-i(n+1/2)\,|\lambda|\,.
\ee
Here $n$ and $l$ are, respectively, the number of the overtone and the angular momentum of the perturbation. The real part of the frequencies is
determined by the angular velocity of the last circular null geodesic $\Omega_m$. The parameter $\lambda$ which appears in the imaginary part is
the Lyapunov exponent which determines the instability timescale of the orbit. For the Lyapunov exponent in \cite{Cardoso} the following formula has been obtained: [the formulas are expressed in term of the chosen anzats for the metric eq. (\ref{metric}), see formulas (40) and (36) in \cite{Cardoso}]

\beq
\lambda =c\,\sqrt{\frac{A_m C''_m-A''_m C_m}{2B_m C_m}}.
\eeq
This expression is reminiscent of the expression (\ref{a}) for the lens parameter $\overline{a}$. (Here and below we restore the speed of light $c$. )
Using the equation for the photon sphere (\ref{Eqm}) and the equation for the impact parameter (\ref{um}), eventually we arrive to the following simple relation
\be
\lambda={c\over u_m \overline{a}}.\label{relat_lambda}
\ee

Another simple observation allows us to relate the coordinate angular velocity with the impact parameter of the lens
\be
\Omega_m= c\,\sqrt{\frac{A_m}{C_m}}={c\over u_m}\label{relat_omega}.
\ee
This relation can be obtained through comparison of the formula for the angular velocity of null geodesics (e.g. formula (37) in the paper of Cardoso et al. \cite{Cardoso}) and formula (\ref{um}). Furthermore, combining (\ref{relat_lambda}) and (\ref{relat_omega}) we find

\be
\overline{a} = \frac{\Omega_m}{\lambda}\label{lambda_omega}.
\ee

A relation between the instability time scale of the orbits and the decrease of the brightness of the images with increasing $n$ can also be obtained.
Let us consider the ratio between the magnifications of two consecutive images. Taking formula (84) from the paper of Bozza \cite{Bozza1} for the magnification in the
limit of large $n$ and formula (\ref{relat_lambda}) we can obtain
\begin{equation}
\ln\left(\frac{\mu_{n+1}}{\mu_n}\right)=-\frac{2\pi}{\overline{a}}=-{2\pi u_m \lambda \over c}
\label{Magnification}
\end{equation}
On the one side, the simple relations (\ref{relat_lambda}) and (\ref{relat_omega}) allow us to give an alternative physical interpretation of the
strong field lens parameter $\overline{a}$ and the impact parameter. On the other side, however, these relations give us a possible method to measure
the real and imaginary parts of the QNM frequencies of different objects since there is good hope that the strong field lens parameters could be
determined in future experiments \cite{Bozza_review, Bozza2}. Equations (\ref{relat_lambda}) and (\ref{relat_omega}) allow us to relate $\lambda$ and $\Omega_m$ to
observable quantities, in particular, the magnitudes of the images, the distance between the observer and the lens $D_{OL}$, and the angular position
of the image that is closest to the black hole $ \theta_{\infty}$ (minimal impact angle). The lens parameter $\overline{a}$ can be determined from the
flux ratio $\tilde{r}$
\be
\overline{a}=\frac{2\pi}{\ln \tilde{r}}\label{a_r},
\ee
defined as
\be
\tilde{r}=\frac{\mu_1}{\sum\limits_{n=2}^\infty \mu_n}.
\ee
Here $\mu_1$ and $\mu_n$ are, respectively, the magnifications of the first image and the $n$-th image. The difference between the magnitude of the outermost relativistic image and magnitude of the sum of all the other relativistic images $r_m$ is related to
the flux ratio $\tilde{r}$ in the following way  $r_m=2.5\, \rm Log  \,\,\tilde{\emph{r}} $. The impact parameter can be obtained from
\begin{equation}
u_m=D_{OL} \theta_{\infty}.
\end{equation}
So, expressed in term of the observables
\be
\lambda={c \ln{\tilde{r}}\over2 \pi D_{OL} \theta_{\infty}}.
\ee
\be
\Omega_m={c\over D_{OL} \theta_{\infty}}.
\ee
Gravitational lensing in the strong field regime is also a candidate for a model independent method for the measurement of the distance to the
observed black holes and other massive objects which act as gravitational lenses. The method presented in
\cite{Bozza2} is based on the measurement of time delays between consecutive relativistic images. According to that method the distance
between the observer and the lens $D_{OL}$ can be determined from the ratio
\be
{\triangle T_{2,1}\over \theta_{\infty}} = 2\pi{D_{OL}\over c},
\ee
where $\triangle T_{2,1}$ is the time delay between the emergence of the first and the second relativistic images and $c$ is the speed of light.
Expressing $D_{OL}$ from the above formula and substituting it back in (\ref{relat_lambda}) and (\ref{relat_omega}) we obtain
\be
\lambda={\ln{\tilde{r}}\over\triangle T_{2,1}}\label{relat_lambda_1}
\ee
and
\be
\Omega_m={2 \pi\over \triangle T_{2,1}}\label{relat_omega_1}.
\ee
The combination of Eqs. (\ref{relat_lambda_1}) and (\ref{relat_omega_1}), or alternatively (\ref{lambda_omega}) and (\ref{a_r}), gives a simple relation between the angular velocity of the last circular null geodesic $\Omega_m$ and the Lyaounov exponent $\lambda$
\be
\Omega_m={2 \pi\over \ln{\tilde{r}}}\lambda.
\ee
Several objects in
different galaxies for which it seems likely that the time delays could be measured in the near future are enlisted in \cite{Bozza2}.

At the end, let us discuss the possible application of the relations between the parameters of the gravitational lens in the strong-deflection regime and the quasinormal modes of static, spherically symmetric black holes in the eikonal approximation presented in the current note. One of the considerable difficulties in future observations of gravitational waves would be the localization of the sources.  The Laser Interferometer Space Antenna (LISA), for example, is an all-sky monitor. According to the estimations, if we try to localize the hosting galaxy of massive black hole(s)  which is a powerful source of gravitational waves the typical LISA error box would contain several hundreds of galaxy clusters which means more than $10^5$ galaxies \cite{Vecchio}. Then, the patch of the sky restricted by LISA has to be additionally examined  with optical or radio telescopes for more precise localization of the source. In that case, the possibility to obtain the characteristic frequencies of emission of gravitational waves of the observed objects through gravitational lensing would be indispensable. Another possible application of the found relations is that if a black hole acts simultaneously as a gravitational lens
and a source of gravitational waves the data from the optical and radio
observations could tell us what frequencies of the gravitational waves
(namely the QNMs) to expect.
%%%%%%%%%%%%%%%%%%%%%%%%%%%%%%%%%%%%%%%%%%%%%%%%%%%%%%%%%%%%%%%%%%%%%%%%%%%%%%%

\section*{Acknowledgements}
This work was partially supported by the Bulgarian National Science Fund under Grants No DO 02-257, No VUF-201/06 and by Sofia University Research Fund
under Grants No 074/2009 and No 101/2010.

%
%%%%%%%%%%%%%%%%%%%%%%%%%%%%%%%%%%%%%%%%%%%%%%%%%%%%%%%%%%%%%%%%%%%%%%%%%%%%%%%

%%%%%%%%%%%%%%%%%%%%%%%%%%%%%%%%%%%%%%%%%%%%%%%%%%%%%%%%%%%%%%%%%%%%%%%%%%%%%%%

\end{document}